# Photoemission Fingerprints for Structural Identification of Titanium Dioxide Surfaces.

*Patrizia Borghetti,*[,1] *Elisa Meriggio,*[1,2] *Gwenaëlle Rousse,*[3] *Gregory Cabailh,*[1] *Rémi Lazzari,*[1] *and Jacques Jupille*[1]

[1] Sorbonne Universités, UPMC Univ Paris 06, CNRS-UMR 7588, Institut des NanoSciences de Paris, F-75005, Paris, France

[2] Dipartimento di Fisica, Università di Genova, Via Dodecaneso 33, 16146 Genova, Italy.

[3] Chimie du Solide et de l'Energie, Collège de France, Sorbonne Universités, UPMC Univ Paris 06, 11, Place Marcelin Berthelot, 75231 Paris, France

**AUTHOR INFORMATION**

**Corresponding Author**

*borghetti@insp.jussieu.fr, Tel.: +33 (0)144274650

Abstract

The wealth of properties of titanium dioxide relies on its various polymorphs and on their mixtures coupled with a sensitivity to crystallographic orientations. It is therefore pivotal to set out methods that allow surface structural identification. We demonstrate herein the ability of photoemission spectroscopy to provide Ti LMV (V = valence) Auger templates to quantitatively analyze $TiO_2$ polymorphs. The Ti LMV decay reflects Ti 4sp-O 2p hybridizations that are intrinsic properties of $TiO_2$ phases and orientations. Ti LMV templates collected on rutile (110), anatase (101), and (100) single crystals allow for the quantitative analysis of mixed nanosized powders, which bridges the gap between surfaces of reference and complex materials. As a test bed, the anatase/rutile P25 is studied both as received and during the anatase-to-rutile transformation upon annealing. The agreement with X-ray diffraction measurements proves the reliability of the Auger analysis and highlights its ability to detect surface orientations.

Introduction

Titanium dioxide (TiO$_2$) triggers an always increasing interest for its innumerous involvements in daily life applications, advanced devices and promising prospects in photochemistry. Among its eight polymorphs,[1] research efforts mainly focus on the most common rutile and anatase phases. Their respective efficiency is still vividly debated since, when comparison is made with nanoparticles (NPs) of similar sizes, rutile may show equal or even higher activity than anatase.[2] However, mixtures of phases that result from synthesis are often of high relevance for application. The archetypal anatase/rutile (80:20) P25 powder[3,4] shows optimal performance as photocatalyst with respect to individual polymorphs.[5,6,7] Beyond the specificity of phases, the chemical activity of titania has a strong orientation-dependent character,[8,9,10] as exemplified by the variety of low-coordinated sites exhibited by P25 particles.[10]

Both phase- and orientation-dependent activity makes crucial the knowledge of the ultimate surface and points to the need of methods with higher surface-sensitivity than X-ray diffraction (XRD)[11,12] and transmission electron microscopy (TEM).[11,13,14] This is all the more true for titania thin films which may escape the sensitivity of XRD. Aside indirect insights gained by selective adsorption,[10] direct chemical analysis is required. Photoemission spectroscopy is the central technique in this respect, as it is currently used to probe defect states, band bending and adsorbates on titania single crystals.[2,15,16] However, the analysis of core-level lines is not able to distinguish the different titania polymorphs. The only differences that were evidenced by photoemission between these polymorphs and surfaces come from the valence bands (VBs). It has been shown by resonant photoemission that the extent of Ti-O hybridization, which marks the VB profile, strongly depends on the polymorph as well as the surface orientation.[17,18] The drawback of the approach is that VBs are inherently difficult to interpret because they combine contributions from both anions and cations. In addition, spectra may involve spurious signals from surface contaminants and, in the case of nanostructured materials, of supporting substrates. The approach is therefore irrelevant for the current use. An alternative is to track Ti Auger transitions involving Ti valence states such as the Ti LMV decay. As previously elucidated by resonant photoemission,[19,20,21] the profile of this transition reflects the VB in such a way that, even if it is significantly perturbed upon reduction,[22] it potentially carries specificities useful to distinguish titanium oxide polymorphs.

The present work investigates this point. Ti LMV Auger spectra of the three more common orientations of TiO$_2$, rutile (110) and anatase (101) and (001) (labeled R(110), A(101) and A(001)) are recorded by laboratory X-ray photoemission spectroscopy in relation with changes in VB spectra.[17,18,23,24] Notably, these orientations correspond to equilibrium shapes (Wulff construction) of anatase, a truncated bipyramid dominated by lateral (101) facets and (001) terminated facets,[8,25-28] and of rutile, mainly (110) facets.[25,26] Similar shapes appear in nature[25,26] and in synthetic powders including P25.[9,10] In a second part, reference Ti LMV spectra are used to analyze phases and orientations involved in (i) bipyramidal anatase NPs (close to Wulff shape) prepared in hydrothermal conditions (labeled HT)[10,29] and P25 NPs (ii) as received and (iii) after the anatase-to-rutile transformation via annealing.[13,30,31] To support the XPS findings, XRD is performed in parallel allowing for the comparison of anatase-to-rutile content ratio.

## Ti LMV spectra of TiO$_2$ single crystals

Since the purpose of the letter is to highlight reference Ti LMV profiles, a detailed presentation of the Ti LMV Auger transition in titanium dioxide is required. The transition is schematically represented in Figure 1a.

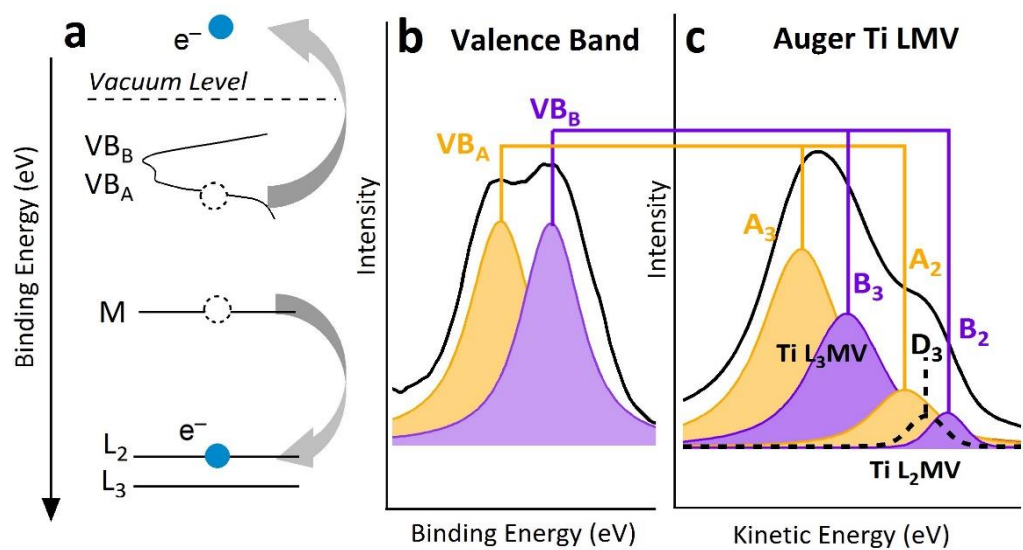

*Figure 1.* Valence band and Ti LMV spectra: (a) schematic representation of the Ti LMV decay (see text for description); (b) valence band of rutile (110) that is decomposed in two indicative components VB$_A$ and VB$_B$; (c) Ti L$_3$MV and Ti L$_2$MV lineshapes recorded on rutile (110); they are decomposed in two components each, A$_2$-B$_2$ and A$_3$-B$_3$ (par. 2 of SI, Figure S3), on the basis of the decomposition of the VB; finally, the component D$_3$ that stems from the presence of band gap states (BGS) is shown.

Following the creation of a hole in a L level, a M electron decaying into that hole causes the emission of a valence band electron (Figure 1a). To illustrate the case, the VB and Ti LMV spectra of the rutile (110) surface are shown in Figure 1b and 1c, respectively. The VB that exhibits the bilobbed profile expected in off-resonance conditions[18,32,33] can be represented by two components labelled VB$_A$ and VB$_B$ (Figure 1b). The Ti LMV transition involves Ti L$_3$MV and Ti L$_2$MV decays (Figure 1c). [19,21] The former stems from L$_3$ holes created either by incident photons or via the fast Coster-Kronig L$_2$L$_3$V relaxation of L$_2$ holes.[19,21] It contributes to the more intense lower kinetic energy (KE) part of the Ti LMV decay (KE ~405-417 eV), while the higher KE part (~417-425 eV) mainly involves Ti L$_2$MV contributions.[19-21] The hierarchy of the kinetic energies follows directly the scheme of the Auger transition (Figure 1a). Reflecting the VB profile (Figure 1b), the Ti L$_3$MV and Ti L$_2$MV lineshapes can be decomposed in two components each, A$_2$-B$_2$ and A$_3$-B$_3$, respectively (Figure 1c), associated with the two main VB features, VB$_A$ and VB$_B$[19,21], with a L$_2$L$_3$ spin-orbit splitting of 5.8 eV[34] (for further details on the decomposition see par. 2 of SI, Figure S3). Finally, due to the partial reduction of the oxide via the formation of O vacancies and interstitial Ti ions, a band-gap state (BGS) associated to the filling of Ti 3d states by excess electrons is expected at a binding energy (BE) ~ 0.8 eV below the Fermi level.[15] The presence of reduced Ti$^{3+}$ was also evidenced in Ti 2p spectra, with a higher concentration of defects in rutile than in anatase (par. 1 of SI, Figure S1). Although the corresponding

BGS was too weak to be detected in the valence band at hν=1486.6 eV, the decomposition of Ti LMV spectra reveals a contribution labelled $D_3$ that is shifted by 7.0 ± 0.2 eV to higher KE relative to $A_3$ (Figure S3). It is assigned to Ti $L_3$MV(BGS) in which the valence level is the BGS.[19] Ti LMV spectra recorded on anatase (101) and (001) were decomposed in a similar way as that of rutile (110). Reproducible $A_3$-$B_3$ ($A_2$-$B_2$) and $A_3$-$D_3$ energy separations validate the reliability of the Ti LMV decomposition (Table S1) which, despite its phenomenological character, reflects the electronic structure (via VB). To avoid unwanted effects due to defects (par. 2 of SI, Figure S4-S5), any forthcoming Ti LMV comparisons will only rely on the lower KE part of the spectra (~405-417 eV).

Spectra recorded on rutile (110) and anatase (101) and (001) – labelled R(110), A(101) and A(001), respectively - are now compared in pair via difference spectra (Figure 2).

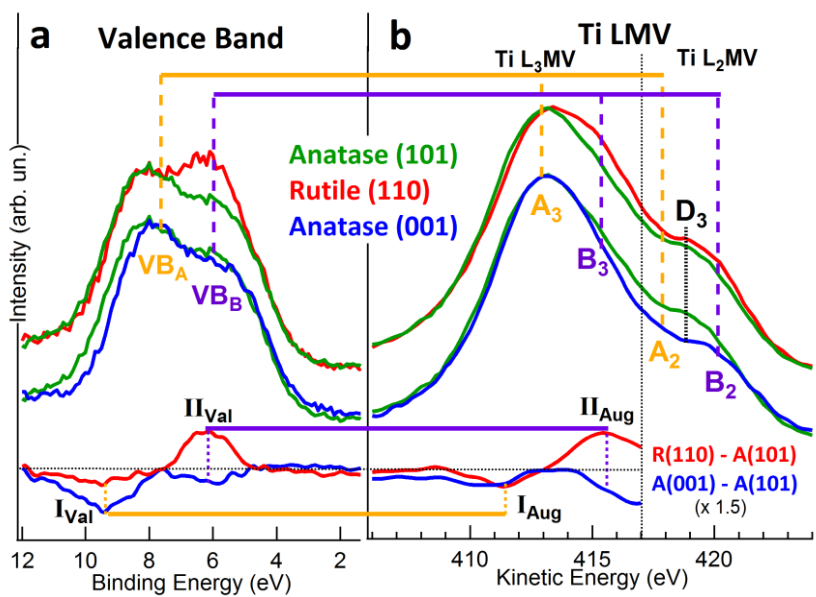

*Figure 2. Top: Comparison in pair of XPS spectra recorded on rutile (110) (red), anatase (101) (green) and (001) (blue) surfaces: (a) VB spectra; (b) Ti LMV Auger spectra. Bottom: difference spectra R(110) - A(101) (red) and A(001) - A(101) (blue). Labels $VB_A$, $VB_B$, $A_2$, $B_2$, $A_3$, $B_3$ and $D_3$ are those of Figure 1 (for further details on the decomposition see par. 2 of SI, Figure S3). VBs and Ti LMV spectra are normalized to $VB_A$ and $A_3$ intensities, respectively. Features $I_{Val}$, $II_{Val}$, $I_{Aug}$, $II_{Aug}$ are discussed in the text. Violet and yellow arrows connect VB features with the corresponding Ti LMV transitions in which they are involved.*

In both rutile and anatase, the molecular-orbital bonding structure associated to the Ti-O octahedral coordination points to Ti 3d-O 2p dominant hybridization with a variable Ti 4sp-O 2p admixture.[23,35,36] Therefore, VB difference spectra (Figure 2a, bottom) reveal marked specificities that originate from variations in the Ti density of state and are reproduced in Ti LMV difference spectra (Figure 2b, bottom). In particular, two features are observed, one in the left part and one in the right part of both VB and Ti LMV difference spectra. First, the higher BE parts of A(001)-A(101) and R(110)-A(101) VB difference spectra are characterized by dips at ~ 9 eV BE (Figure 2a). In the case of A(001)-A(101), the intense dip arises from the narrowness of the $VB_A$ component of A(001) with respect to A(101) (Figure 2a, $I_{val}$ blue). Resonant photoemission partly accounts for that difference by showing that Ti 4sp states contribute almost nothing to the VB$_A$ feature of A(001).[17] In the case of R(110)-A(101), the dip I$_{val}$ red (Figure 2a, bottom) originates from the leftward shift (~0.1 eV) of VB$_A$ component in the A(101) spectrum. The blue and red I$_{val}$ dips match the blue and red I$_{Aug}$ dips at KE ~411 eV in Ti L$_3$MV difference spectra (Figure 2b, bottom). As in the case of VB spectra, the negative dip of R(110)-A(101) spectrum originates from the leftward shift (0.15 eV) of the A$_3$ component in the A(101) spectrum.

Next, the lower BE part of the A(001)-A(101) and R(110)-A(101) VB difference spectra show a slight dip (Figure 2a, II$_{val}$ blue) and an intense peak (II$_{val}$ red), respectively, at BE~ 6 eV. They correspond to the dip (II$_{Aug}$ blue) and the peak (II$_{Aug}$ red) observed at KE ~415-416 eV in the A(001)-A(101) and R(110)-A(101) Ti L$_3$MV difference spectra, respectively. These features mainly arise from the variable Ti 4sp character of the VB$_B$ density of states which increases in the order A(001)<A(101)<R(110).[18,24,37] The strong Ti 4sp-O 2p hybridization evidenced by resonant[18,37] and site specific XPS[24] in R(110) is suggested to stem from the different arrangement of Ti-centered octahedra; rutile is denser that anatase enhancing the O 2p-Ti 4sp overlap.[23,33,37] Moreover, the dip in the A(001)-A(101) difference (Figure 1a, II$_{val}$ blue and Figure 1b, II$_{Aug}$ blue) is consistent with the decrease in Ti 4sp admixture on anatase (001) observed by resonant photoemission.[17]

**Ti LMV spectra of TiO$_2$ nanoparticles**

The Ti LMV references recorded on single crystals are used to determine phases and orientations of the P25 and the hydrothermal HT[10,29] powders deposited on gold substrates. The XRD-determined anatase:rutile content is 86:14 (±3%) in P25 with coherent crystalline domains of 23±4 nm and 28±8 nm, respectively (Figure S6b). HT powder involves pure anatase NPs with domains of 24±4 nm (Figure S6a). In Figures 3a and 3b, Ti LMV spectra are normalized to the intensity of peak A$_3$ and superimposed to the single-crystal spectra. Visual inspection shows that P25 and HT mainly exhibit A(101) facets, in agreement with CO adsorption measurements.[10]

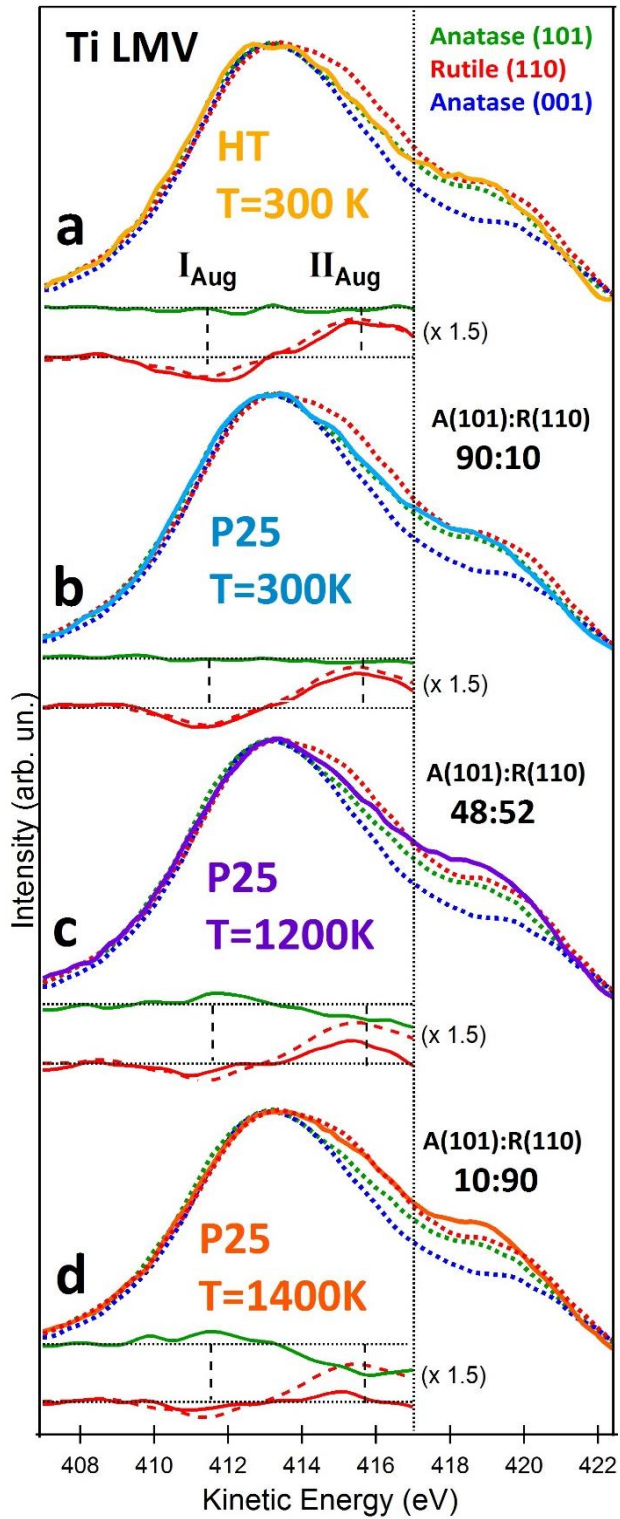

*Figure 3. Comparison of Ti LMV spectra of (a) HT (yellow line) and of P25 (b) at T=300 K (blue line), (c) after annealing at T=1200 K (violet line) and (d) T=1400 K (orange line) to R(110), A(101) and A(001) spectra from Figure 1b (red, green and blue dotted lines, respectively). Difference spectra with respect to A(101) and R(110) (green and red solid lines, respectively) along with the difference spectrum R(110)-A(101) (red dashed line) are also shown in the bottom of each panel. The content ratio of P25 obtained from Auger analysis is given in panels b,c,d.*

Quantitatively, difference spectra are compared on the basis of the reliance value $x^2$ (par. 4-5 of SI). An almost zero difference spectrum is obtained when the spectrum of anatase (101) is subtracted to the spectrum of each powder (solid line of Figure 3a, 3b), with $x^2$=2.6 for HT and $x^2$=1.6 for P25. The correspondence of P25 and HT to R(110) is much weaker, as indicated by the $x^2$ of the relative difference spectra, which for both powders is higher than 22. It can be also observed that in the difference spectra between R(110) and powders features $I_{Aug}$ and $II_{Aug}$ occur at the same positions and intensities as in R(110)-A(101) (Figure 2b).

The relative concentration of P25 polymorphs and facets has also been quantified through a linear combination of R(110), A(101) and A(001) spectra. The fit is evaluated in the energy range ~405-417 eV, *i.e.* the part of the spectrum not including the Ti L$_3$MV(BGS) decay (Linear combination spectra, test and estimation of the error are presented in SI, par. 5, Figures S7-S8). This leads to an anatase-to-rutile ratio of 90:10 (±10%) (Table 1), without any contribution from A(001). The $x^2$ difference between P25 and the present linear combination shows a slight improvement with respect to the P25–A(101) difference (1.4 vs 1.6) (Figure S7).

| Sample | Method | Anatase% | Rutile% | Error% |
|---|---|---|---|---|
| P25 | XRD | 86 | 14 | ±3 |
| | XRD correc | 88 | 12 | |
| | Ti LMV | 90 (101) | 10 (110) | ±10 |
| P25 T=1200 K | XRD | 3 | 97 | ±3 |
| | XRD correc | 6 | 94 | |
| | Ti LMV | 48 (101) | 52 (110) | ±5 |
| P25 T=1400 K | XRD | 3 | 97 | ±3 |
| | XRD correc | 6 | 94 | |
| | Ti LMV | 10 (101) | 90 (110) | ±10 |

***Table 1.*** *Composition of P25 powders as determined by XRD and Ti LMV analysis.*

The comparison of Ti LMV analysis to XRD requires a correction of XRD ratios by the ratio of the particle radii R to account for the fact that Ti LMV analysis probes powder surfaces while XRD characterizes the bulk (XRD correc, Table 1):

$$\frac{[anatase]}{[rutile]}_{(Auger)} = \frac{[anatase]_{XRD}}{[anatase]_{XRD}+[rutile]_{XRD}\cdot\frac{R_{anatase}}{R_{rutile}}}$$

Assuming that R is given by the domain size obtained by XRD, a 88:12 ratio is found, in good agreement with the estimate of 90:10 obtained by Ti LMV analysis (Table 1). Beyond this matching on polymorph content, only surface-sensitive Ti LMV analysis allows determining the dominant facet of anatase P25 nanoparticles, i.e. the (101).

Finally, the anatase-to-rutile transformation of P25[13,30,31] was performed in UHV by annealing the support of the powder at 1200 K and 1400 K (Figure 3c,d). *Ex situ* XRD (Figure S6c,d) evidences an almost complete anatase-to-rutile transformation at both temperatures with a rutile content of 97±3% (Table 1). Domain sizes of anatase and rutile particles were 19±8 nm and 51±4 nm, respectively, after annealing at 1200 K, and 23±8 nm and 63±4 nm after annealing at 1400 K. The increase in size and the observed hierarchy (rutile>anatase) correspond to expectation.[13,30,38] Regarding the Ti LMV profile, the anatase-to-rutile transformation is revealed by the increase in the $B_3/A_3$ area ratio and the shift of $A_3$ towards higher KE (Figure 3c,d). The increase in intensity at KE ~419 eV due to the reduction of the sample (component $D_3$) does not affect the low KE range ~405-417 eV on which the quantitative analysis is based (Figure S5). The difference between the P25 and R(110) spectra further confirms the transformation (bottom spectra of Figure 3c,d): while $\chi^2$=22 in the case of the as-received P25, $\chi^2$ decreases to 9.3 and 2.0 after annealing at T=1200 K and T=1400 K, respectively.

After annealing at 1400 K, the Ti LMV-based measurements nicely agree with the corrected XRD values, with anatase:rutile ratios of 10:90 and 6:94, respectively (Table 1), demonstrating the reliability of the Ti LMV analysis. In contrast, after annealing at 1200 K, the anatase-to-rutile ratio of 48:52 found via Ti LMV seems to contradict the XRD ratio of 7:93 (Table 1). This apparent discrepancy is explained by the occurrence of a temperature gradient between the metallic support and the extreme surface of the titania powder. Bulk XRD indicates that the two-hours-long 1200 K annealing allows for an almost complete transformation, while the surface-sensitive XPS probes the untransformed fraction of the extreme surface which is indicative of a vertical temperature gradient.

To summarize, Ti LMV Auger reference spectra of rutile (110), anatase (101) and (001) crystals are singled out by taking advantage of the valence level involved in the Auger line. They mostly differ from the way Ti 4sp states hybridize with O 2p states in various phases and orientations. From Auger templates, powder facets contents are quantified, in particular for P25 in the as-received form and during the anatase-to-rutile transformation upon annealing. Comparison to X-ray diffraction demonstrates the solidity of the Auger analysis and discloses its general ability to detect surface orientations. Noteworthily, the distinct features of Ti LMV spectra are adequately evidenced by a nonmonochromatic X-ray lamp, making the present method suitable for home-laboratory analysis. The present quantification method opens interesting perspectives for the characterization of metal-oxide phases and orientations, in particular for thin films which may escape to the sensitivity of XRD measurements.

**Experimental Methods**

The anatase (5x5 mm$^2$) natural single crystals of (101) and (001) orientations and the synthetic (10x10 mm$^2$) rutile (110) single crystal were supplied by Mateck GmbH. The three samples were prepared by Ar$^+$ ion sputtering cycles of 1 keV for 10 min followed by 20 min annealing at 850 K and 1000 K for anatase and rutile, respectively. The absence of surface contaminants was checked at the level of XPS sensitivity (a few % of monolayer) while the surface crystallinity was indicated by sharp LEED patterns (1x1) for rutile (110) and anatase (101) crystals and (1x4) for the anatase (001) sample. The sputtered sample of rutile (110) crystal was obtained by one Ar$^+$ ion sputtering cycles of 800 eV without subsequent thermal annealing (Figure S4). In this case, no LEED pattern was obtained.

The P25 nanopowder was supplied by Evonik Industries. The HT nanoparticles were prepared in hydrothermal conditions (HT) following the procedure described in Refs.[10] and [29], respectively. These nanoparticles are of the UT001 type produced at the Department of Chemistry of the University of Torino in the frame of the FP7 EU project SETNanoMetro (www.setnanometro.eu). For XPS measurements and annealing, we used an alumina $Al_2O_3$(0001) supporting substrate covered by 1 nm of NiCr alloy and a 200 nm-thick gold layer. Thermal annealing at T=600 K allowed to partially remove surface contaminants from the powders. We verified that the Ti LMV spectra of the treated samples (Figures 2-3 of the article), apart from improving the signal-to-noise ratio, were identical to the spectra of the as-received powders. The anatase-to-rutile transformation of P25 was obtained by electron-bombarding in vacuum the back side of the Mo holder on which the P25 samples were mounted (powders/Au/NiCr/$Al_2O_3$). The transformation has been performed at two temperatures, T = 1200 K and T = 1400 K, both measured on the Mo plate by an optical pyrometer and maintained for 2 hours.

XPS spectra (Ti LMV and VB spectra) were obtained with a nonmonochromatic Al K$\alpha$ source (photon energy of h$\nu$=1486.7 eV) and a hemispherical analyzer Phoibos 100 with a pass energy of 20 eV (valence band spectra) and 50 eV (Ti LMV spectra) at normal emission. The Ti LMV and VB difference spectra are smoothed by applying a 1-pass binomial filter.[39] A linear background was subtracted to all the Ti LMV spectra in order to equal the intensity at KEs of 401 eV and 430 eV (for further details see Supporting Information).

**Supporting Information**

XPS Ti $2p_{3/2}$ spectra of rutile (110) and anatase (101) (par. 1, Figure S1), normalization and fitting of Ti LMV spectra of single crystals (par. 2, Figures S2-S3 and Table S1), effects of Ar$^+$ sputtering and annealing at high temperatures on the Ti LMV lineshape (par. 2, Figures S4-S5), X-ray powder diffraction patterns (par. 3, Figure S6), definition of the reliance factor $\chi^2$ (par. 4, Equations S1-S2), linear combination spectra fitting the Ti LMV spectra of powders (par. 5, Figure S7-S8).
This material is available free of charge via the Internet at http://pubs.acs.org.

**AUTHOR INFORMATION**

**Notes**

The authors declare no competing financial interests.

**ACKNOWLEDGMENT**

This work was supported by the FP7 SETNanoMetro project (no. 604577). P. B. acknowledges financial support from the European Research Council (ERC) under Horizon 2020 research and innovation programme (Grant agreement No 658056). Authors acknowledge V. Maurino and F. Pellegrino (Chemistry Department of the University of Torino, Italy) for the synthesis of $TiO_2$-HT powders.

**REFERENCES**

(1) Yang, Z.; Choi, D.; Kerisit, S.; Rosso, K. M.; Wang, D.; Zhang, J.; Graff, G.; Liu, J. Nanostructures and Lithium Electrochemical Reactivity of Lithium Titanites and Titanium Oxides: A Review. *J. Power Sources* **2009**, *192*, 588-598.

(2) Henderson, M. A. A surface science perspective on $TiO_2$ photocatalysis. *Surf. Sci. Rep.* **2011**, *66*, 185-297.

(3) Hurum, D. C.; Agrios, A. G.; Gray, K. A.; Rajh, T.; Thurnauer, M. C. Explaining the Enhanced Photocatalytic Activity of Degussa P25 Mixed-Phase $TiO_2$ Using EPR. *J. Phys. Chem. B* **2003**, *107*, 4545-4549.

(4) Ohtani, B.; Prieto-Mahaney, O. O.; Li, D.; Abe, R. What is Degussa (Evonik) P25? Crystalline Composition Analysis, Reconstruction from Isolated Pure Particles and Photocatalytic Activity Test. *J. Photochem. Photobiol. A* **2010**, *216*, 179-182.

(5) Tsukamoto, D.; Shiraishi, Y.; Sugano, Y.; Ichikawa, S.; Tanaka, S.; Hirai, T. Gold Nanoparticles Located at the Interface of Anatase/Rutile $TiO_2$ Particles as Active Plasmonic Photocatalysts for Aerobic Oxidation. *J. Am. Chem. Soc.* **2012**, *134*, 6309-6315.

(6) Liu, N.; Schneider, C.; Freitag, D.; Venkatesan, U.; Marthala, V. R. R.; Hartmann, M.; Winter, B.; Spiecker, E.; Osvet, A.; Zolnhofer, E. M.; Meyer, K.; Nakajima, T.; Zhou, X.; Schmuki, P. Hydrogenated Anatase: Strong Photocatalytic Dihydrogen Evolution without the Use of a Co-Catalyst. *Angew. Chem. Int. Ed.* **2014**, *126*, 14425-14429.

(7) Priebe, J. B.; Radnik, J.; Lennox, A. J. J.; Pohl, M.-M.; Karnahl, M.; Hollmann, D.; Grabow, K.; Bentrup, U.; Junge, H.; Beller, M.; Brückner, A. Solar Hydrogen Production by Plasmonic Au–$TiO_2$ Catalysts: Impact of Synthesis Protocol and $TiO_2$ Phase on Charge Transfer Efficiency and $H_2$ Evolution Rates. *ACS Catal.* **2015**, *5*, 2137-2148.

(8) Barnard, A. S.; Zapol, P. Effects of Particle Morphology and Surface Hydrogenation on the Phase Stability of $TiO_2$. *Phys. Rev. B* **2004**, *70*, 235403.

(9) Yang, H. G.; Sun, C. H.; Qiao, S. Z.; Zou, J.; Liu, G.; Smith, S. C.; Cheng, H. M.; Liu, G. Q. Anatase $TiO_2$ Single Crystals with a Large Percentage of Reactive Facets. *Nature* **2008**, *453*, 638-641.

(10) Deiana, C.; Minella, M.; Tabacchi, G.; Maurino, V.; Fois, E.; Martra, G. Shape-controlled $TiO_2$ Nanoparticles and $TiO_2$ P25 Interacting with CO and $H_2O_2$ Molecular Probes: A Synergic Approach for Surface Structure Recognition and Physico-Chemical Understanding. *Phys. Chem. Chem. Phys.* **2013**, *15*, 307-315.

(11) Chu, L.; Qin, Z.; Yang, J.; Li, X. Anatase TiO$_2$ Nanoparticles with Exposed {001} Facets for Efficient Dye-Sensitized Solar Cells. *Sci. Rep.* **2015**, *5*, 12143.
(12) Sakurai, K.; Mizusawa, M. X-Ray Diffraction Imaging of Anatase and Rutile. *Anal. Chem.* **2010**, *82*, 3519-3522.
(13) Gouma, P. I.; Mills, M. J. Anatase-to-Rutile Transformation in Titania Powders. *J. Am. Ceram. Soc.* **2001**, *84*, 619-622.
(14) Ohno, T.; Sarukawa, K.; Tokieda, K.; Matsumura, M. Morphology of a TiO$_2$ Photocatalyst (Degussa, P-25) Consisting of Anatase and Rutile Crystalline Phases. *J. Catal.* **2001**, *203*, 82-86.
(15) Diebold, U. The Surface Science of Titanium Dioxide. *Surf. Sci. Rep.* **2003**, *48*, 53-229.
(16) Zhang, Z.; Yates, J. T. Jr. Band Bending in Semiconductors: Chemical and Physical Consequences at Surfaces and Interfaces. *Chem. Rev.* **2012**, 112, 5520-5551.
(17) Thomas, A. G.; Flavell, W. R.; Kumarasinghe, A. R.; Mallick, A. K.; Tsoutsou, D.; Smith, G. C.; Stockbauer, R.; Patel, S.; Grätzel, M.; Hengerer, R. Resonant Photoemission of Anatase TiO$_2$ (101) and (001) Single Crystals. *Phys. Rev. B* **2003**, *67*, 035110.
(18) Thomas, A. G.; Flavell, W. R.; Mallick, A. K.; Kumarasinghe, A. R.; Tsoutsou, D.; Khan, N.; Chatwin, C.; Rayner, S.; Smith, G. C.; Stockbauer, R. L.; et al. Comparison of the Electronic Structure of Anatase and Rutile TiO$_2$ Single-Crystal Surfaces Using Resonant Photoemission and X-Ray Absorption Spectroscopy. *Phys. Rev. B* **2007**, *75*, 035105.
(19) Le Fèvre, P.; Danger, J.; Magnan, H.; Chandesris, D.; Jupille, J.; Bourgeois, S.; Arrio, M.-A.; Gotter, R.; Verdini, A.; Morgante, A. Stoichiometry-Related Auger Lineshapes in Titanium Oxides: Influence of Valence-Band Profile and of Coster-Kronig Processes. *Phys. Rev. B* **2004**, *69*, 155421.
(20) Jupille, J.; Chandesris, D.; Danger, J.; Le Fèvre, P.; Magnan, H.; Bourgeois, S.; Gotter, R.; Morgante, A. Resonant L$_2$MV and L$_3$MV Auger Transitions in Titanium Dioxide. *Surf. Sci.* **2001**, *482-485*, 453-457.
(21) Danger, J.; Magnan, H.; Chandesris, D.; Le Fèvre, P.; Bourgeois, S.; Jupille, J.; Verdini A.; Gotter, R.; Morgante, A. Intra-Atomic Versus Interatomic Process in Resonant Auger Spectra at the Ti L$_{23}$ Edges in Rutile. *Phys. Rev. B* **2001,** *64*, 045110.
(22) Henrich, V. E.; Dresselhaus, G.; Zeiger, H. J. Surface Defects and the Electronic Structure of SrTiO$_3$ Surfaces. *Phys. Rev. B* **1978,** 17, 4908-4921.
(23) Asahi, R.; Taga, Y.; Mannstadt, W.; Freeman, A. J. Electronic and Optical Properties of Anatase TiO$_2$. *Phys. Rev. B* **2000,** *61*, 7459-7465.
(24) Woicik, J. C.; Nelson, E. J.; Kronik, L.; Jain, M.; Chelikowsky, J. R.; Heskett, D.; Berman, L. E.; Herman, G. S. Hybridization and Bond-Orbital Components in Site-Specific X-Ray Photoelectron Spectra of Rutile TiO$_2$. *Phys. Rev. Lett.* **2002**, *89*, 077401.
(25) Ziolkowski, J. New Method of Calculation of the Surface Enthalpy of Solids. *Surf. Sci.* **1989**, *209*, 536-561.
(26) Ramamorphy, M.; Vanderbilt, D.; King-Smith, R. D. First-Principles Calculations of the Energetics of Stoichiometric TiO$_2$ Surfaces. *Phys. Rev. B* **1994**, *49*, 16721-16727.
(27) Oliver, P. M.; Watson, G. W.; Kelsey, E. T.; Parker, S. C. Atomistic Simulation of the Surface Structure of the TiO$_2$ Polymorphs Rutile and Anatase. *J. Mater. Chem.* **1997**, *7*, 563-568.
(28) Lazzeri, M.; Vittadini, A.; Selloni, A. Structure and Energetics of Stoichiometric TiO$_2$ Anatase Surfaces. *Phys. Rev. B* **2001**, *63*, 155409.

(29) Deiana, C.; Fois, E.; Martra, G.; Narbey, S.; Pellegrino, F.; Tabacchi, G. On the Simple Complexity of Carbon Monoxide on Oxide Surfaces: Facet-Specific Donation and Backdonation Effects Revealed on TiO$_2$ Anatase Nanoparticles. *Chem. Phys. Chem.* **2016**, *17*, 1956-1960.
(30) Gribb, A. A.; Banfield, J. F. Particle Size Effects on Transformation Kinetics and Phase Stability in Nanocrystalline TiO$_2$. *Am. Mineral.* **1997**, *82*, 717-728.
(31) Hanaor, D. A. H.; Sorrell, C. C. Review of the Anatase to Rutile Phase Transformation. *J. Mater. Sci*. **2011**, *46*, 855-874.
(32) Zhang, Z.; Jeng, S.-P.; Henrich, V. E. Cation-Ligand Hybridization for Stoichiometric and Reduced TiO$_2$(110) Surfaces Determined by Resonant Photoemission. *Phys. Rev. B* **1991**, *43*, 12004-12011.
(33) Pfeifer. V.; Erhart, P.; Li, S.; Rachut, K.; Morasch, J.; Brötz, J.; Reckers, P.; Mayer, T.; Rühle, S.; Zaban, A.; et al. Energy Band Alignment Between Anatase and Rutile TiO$_2$. *J. Phys. Chem. Lett.* **2013**, *4*, 4182-4187.
(34) Biesinger, M. C.; Lau, L. W. M.; Gerson, A. R.; Smart, R. St. C. Resolving Surface Chemical States in XPS Analysis of First Row Transition Metals, Oxides and Hydroxides: Sc, Ti, V, Cu and Zn. *Appl. Surf. Sci.* **2010**, *257*, 887-898.
(35) Fisher, D. W. X-Ray Band Spectra and Molecular-Orbital Structure of Rutile TiO$_2$. *Phys. Rev. B* **1972**, *5*, 4219-4226.
(36) Brydson, R.; Sauer, H.; Engel, W.; Thomas, J. M.; Zeitler, E.; Kosugi, N.; Kuroda, H. Electron Energy Loss and X-Ray Absorption Spectroscopy of Rutile and Anatase: A Test of Structural Sensitivity. *J. Phys.: Condens. Matter* **1989**, *1*, 797-812.
(37) Nerlov, J.; Ge, Q.; Møller, P. J. Resonant Photoemission from TiO$_2$(110) Surfaces: Implications on Surface Bonding and Hybridization. *Surf. Sci.* **1996**, *348*, 28-38.
(38) Zhang, H.; Banfield, J. F. Thermodynamic Analysis of Phase Stability of Nanocrystalline Titania. *J. Mat. Chem.* **1998**, *8*, 2073-2076.
(39) Marchand, P.; Marmet, L. Binomial Smoothing Filter: A Way to Avoid Some Pitfalls of Least-Squares Polynomial Smoothing. *Rev. Sci. Instrum.* **1983**, *54,* 1034-1041.